\documentclass[%
reprint,
showkeys,
amsmath,amssymb,
aps, physrev,
prl,
longbibliography
]{revtex4-2}

\usepackage[colorlinks = true,
            linkcolor = blue,
            urlcolor  = blue,
            citecolor =  blue,
            anchorcolor = blue]{hyperref}
\usepackage{graphicx,float}
\usepackage{xcolor}

\usepackage{bm}
\usepackage{dsfont}
\usepackage{stmaryrd}
\usepackage{tikz}
\usetikzlibrary{calc}

\newcommand{\ds}{\mathrm{d}}
\newcommand{\p}{\partial}
\newcommand{\bb}{\mathbb}

\newcommand{\cc}{\mathcal}
\RequirePackage[normalem]{ulem} 
\RequirePackage{color}\definecolor{RED}{rgb}{1,0,0}\definecolor{BLUE}{rgb}{0,0,1} 
\providecommand{\DIFaddbegin}{} 
\providecommand{\DIFaddend}{} 
\providecommand{\DIFdelbegin}{} 
\providecommand{\DIFdelend}{} 
\providecommand{\DIFaddbeginFL}{} 
\providecommand{\DIFaddendFL}{} 
\providecommand{\DIFdelbeginFL}{} 
\providecommand{\DIFdelendFL}{} 
\newcommand{\DIFscaledelfig}{0.5}
\RequirePackage{settobox} 
\RequirePackage{letltxmacro} 
\newsavebox{\DIFdelgraphicsbox} 
\newlength{\DIFdelgraphicswidth} 
\newlength{\DIFdelgraphicsheight} 
\LetLtxMacro{\DIFOincludegraphics}{\includegraphics} 
\newcommand{\DIFaddincludegraphics}[2][]{{\color{blue}\fbox{\DIFOincludegraphics[#1]{#2}}}} 
\newcommand{\DIFdelincludegraphics}[2][]{
\sbox{\DIFdelgraphicsbox}{\DIFOincludegraphics[#1]{#2}}
\settoboxwidth{\DIFdelgraphicswidth}{\DIFdelgraphicsbox} 
\settoboxtotalheight{\DIFdelgraphicsheight}{\DIFdelgraphicsbox} 
\scalebox{\DIFscaledelfig}{
\parbox[b]{\DIFdelgraphicswidth}{\usebox{\DIFdelgraphicsbox}\\[-\baselineskip] \rule{\DIFdelgraphicswidth}{0em}}\llap{\resizebox{\DIFdelgraphicswidth}{\DIFdelgraphicsheight}{
\setlength{\unitlength}{\DIFdelgraphicswidth}
\begin{picture}(1,1)
\thicklines\linethickness{2pt} 
{\color[rgb]{1,0,0}\put(0,0){\framebox(1,1){}}}
{\color[rgb]{1,0,0}\put(0,0){\line( 1,1){1}}}
{\color[rgb]{1,0,0}\put(0,1){\line(1,-1){1}}}
\end{picture}
}\hspace*{3pt}}} 
} 
\LetLtxMacro{\DIFOaddbegin}{\DIFaddbegin} 
\LetLtxMacro{\DIFOaddend}{\DIFaddend} 
\LetLtxMacro{\DIFOdelbegin}{\DIFdelbegin} 
\LetLtxMacro{\DIFOdelend}{\DIFdelend} 
\DeclareRobustCommand{\DIFaddbegin}{\DIFOaddbegin \let\includegraphics\DIFaddincludegraphics} 
\DeclareRobustCommand{\DIFaddend}{\DIFOaddend \let\includegraphics\DIFOincludegraphics} 
\DeclareRobustCommand{\DIFdelbegin}{\DIFOdelbegin \let\includegraphics\DIFdelincludegraphics} 
\DeclareRobustCommand{\DIFdelend}{\DIFOaddend \let\includegraphics\DIFOincludegraphics} 
\LetLtxMacro{\DIFOaddbeginFL}{\DIFaddbeginFL} 
\LetLtxMacro{\DIFOaddendFL}{\DIFaddendFL} 
\LetLtxMacro{\DIFOdelbeginFL}{\DIFdelbeginFL} 
\LetLtxMacro{\DIFOdelendFL}{\DIFdelendFL} 
\DeclareRobustCommand{\DIFaddbeginFL}{\DIFOaddbeginFL \let\includegraphics\DIFaddincludegraphics} 
\DeclareRobustCommand{\DIFaddendFL}{\DIFOaddendFL \let\includegraphics\DIFOincludegraphics} 
\DeclareRobustCommand{\DIFdelbeginFL}{\DIFOdelbeginFL \let\includegraphics\DIFdelincludegraphics} 
\DeclareRobustCommand{\DIFdelendFL}{\DIFOaddendFL \let\includegraphics\DIFOincludegraphics} 
\RequirePackage{listings} 
\RequirePackage{color} 
\lstdefinelanguage{DIFcode}{ 
  moredelim=[il][\color{red}\sout]{\%DIF\ <\ }, 
  moredelim=[il][\color{blue}\uwave]{\%DIF\ >\ } 
} 
\lstdefinestyle{DIFverbatimstyle}{ 
	language=DIFcode, 
	basicstyle=\ttfamily, 
	columns=fullflexible, 
	keepspaces=true 
} 
\lstnewenvironment{DIFverbatim}{\lstset{style=DIFverbatimstyle}}{} 
\lstnewenvironment{DIFverbatim*}{\lstset{style=DIFverbatimstyle,showspaces=true}}{} 

\begin{document}


\preprint{APS/123-QED}

\title{Hidden causal fingerprint of self-consistent transport}

\author{Daniel Yaacoub$^\dagger$}
\email[Corresponding author: ]{yaacoub@laplace.univ-tlse.fr}
\author{Idriss Adjaout$^\star$, St\'ephane Blanco$^\dagger$, Jean-Fran\c cois Cornet$^\star$, J\'er\'emi Dauchet$^\star$, Richard Fournier$^\dagger$, Thomas Vourc'h$^\star$}

\affiliation{
$^\dagger$UPS, CNRS, INPT, LAPLACE UMR CNRS 5213, Universit\'e de Toulouse,\\
118 route de Narbonne, F-31065 Toulouse, Cedex 9, France.\\
$^\star$Universit\'e Clermont Auvergne, Clermont Auvergne INP,\\
CNRS, Institut Pascal, F-63000 Clermont-Ferrand, France.
}

\begin{abstract}
Linear transport is fundamentally causal: 
information propagates from physical sources to observable effects along well-defined histories materialized by paths such as light rays, Brownian trajectories or fluid streamlines. 
Self-consistent transport presents a conceptual paradox: 
because the transport-field depends on the transported state, source-to-observable propagation is obscured, and the direct pathwise causality is no longer explicit.
  Here we show that causality is not lost; it is hidden.
  Information flow is organized through branching histories.
By revealing this previously hidden physical structure of information organization, we recover a measurable source-to-observable causal map.
We demonstrate that a local change in morphology reorganizes entirely the hidden physical structure of self-consistent transport while leaving unchanged the observable state.
\end{abstract}

\maketitle

\begin{figure}[H]
  \includegraphics[trim=7mm 12mm 10mm 1.3cm,clip,width=1\linewidth]{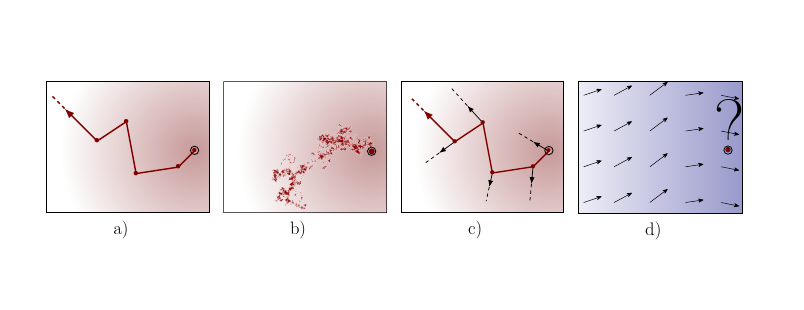}
  \caption{
   \scriptsize{
   a) Light ray backwardly propagating from a probe point to the corresponding physical source \cite{Kakutani_1944,Feynman_1948,Kac_1951,Einstein_1905,Smoluchowski_1906b}.
   b) Brownian trajectory as a history of diffusion \cite{Villefranque_2022,Ibarrart_2025}.
   c) Branching tree as a history of colliding molecules in Boltzmann gas kinetics \cite{Nyffenegger_2024,Terree_2022,Pulvirenti_2018,Kac_1956,McKean_1966,McKean_1967} and reaction-diffusion \cite{Skorokhod_1964,McKean_1975} models.
   d) Hidden causal history underlying self-consistent transport.}
  }
  \label{fig:history}
\end{figure}

\textbf{Introduction.} -- \textit{How physical influences propagate through space and time from a physical source to an observable?} 
From chemical and heat diffusion to light scattering, fluid and plasma dynamics, this propagation admits a direct pathwise reading along rays, characteristics, streamlines or Brownian trajectories (Fig.~\ref{fig:history}) \cite{Kakutani_1944,Feynman_1948,Kac_1951,Onsager_1953,Wiener_1921}. 
This identifies the fundamental histories through which information flows across a system. 
From Einstein's description of diffusion to Feynman's picture of quantum mechanics, this trajectory viewpoint has become a cornerstone of modern  physics \cite{Einstein_1905,Phillips_1923,Budaev_2002,Kac_1974,Zhang_2019,Maire_2006,Lejay_2010,Tessendorf_1987}.
A path then provides more than a representation of the observable: it identifies the history connecting the observation event to its physical source.

This causal picture is exact when the transport field is prescribed independently of the transported state of the system. 
However, many interacting systems are intrinsically self-consistent: 
the transport field itself depends on the evolving state that it transports. 
This occurs across a broad range of field-mediated dynamics, from plasma, globular-cluster, and two-phase dynamics (Poisson--Vlasov and Enskog--Vlasov kinetic models), to fluid mechanics, electrochemical transport and active matter (Navier--Stokes, Poisson--Nernst--Planck and Keller--Segel macroscopic models). 
The propagation mechanism is no longer an imposed background, but is part of the dynamics itself and the transport field depends on all the transported states across space and time. 
Effective descriptions correctly predict observables, but they do not provide a direct reconstruction of how information propagates from physical sources to those observables.

\begin{figure*}[t!] 
  \centering
  \includegraphics[clip,width=1\linewidth]{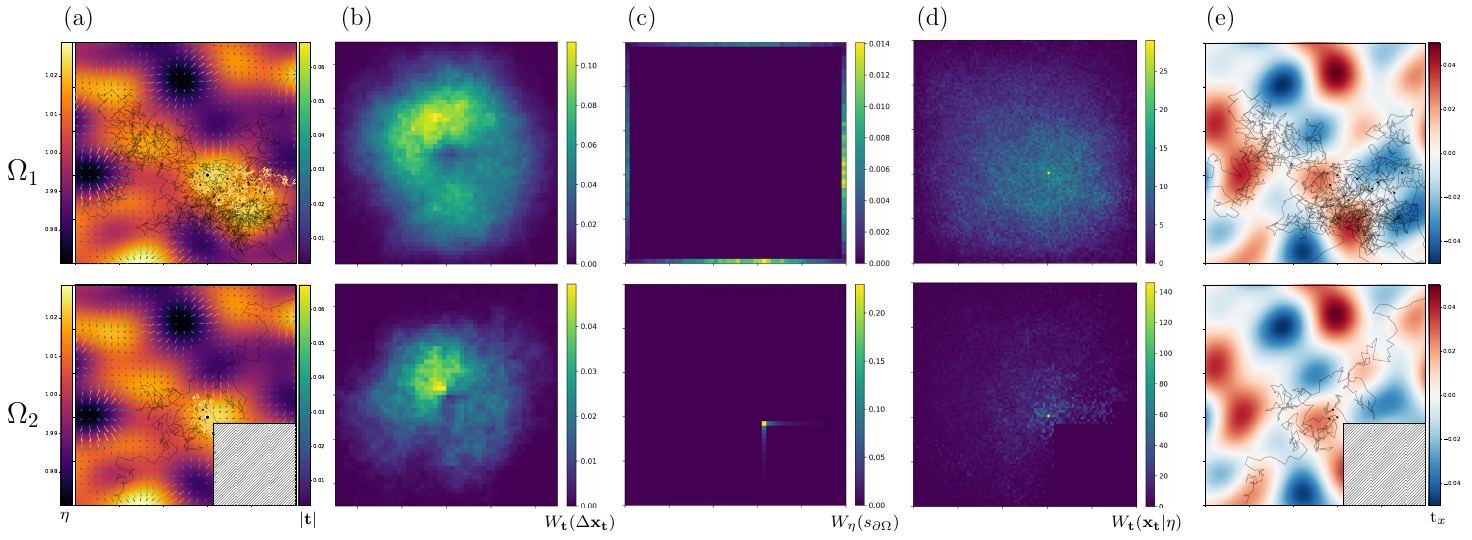}
  \caption{
   \scriptsize{
Comparison of $\Omega_1$ (top, reference) and $\Omega_2$ (bottom, locally modified), which yield the same observable state. The panels resolve complementary levels of the same source-to-probe causal fingerprint.
(a) Observable history (yellow) joint to local transport-field realizations (black).
The colormap depicts the observable field solution while the vector map depicts the transport-field (see Appendix).
(b) Contribution-weighted spatial distribution of the hidden transport-field histories relative to their final boundary hit. Each point sampled along a hidden transport history inherits the contribution weight of its parent observable history, revealing which regions of the transport environment participate most strongly in carrying the probe response.
(c) Contribution-weighted boundary-hit distribution.
It resolves how the observable response is distributed over its source locations, complementing the purely geometric accessibility measured by the hitting distribution.
(d) Contribution-weighted transport-history positions conditioned on $1.04<\eta<1.12$. These conditional maps identify the regions of the hidden transport environment associated with histories carrying a selected range of observable values.
(e) Sampled $x$-component transport-field realizations along the hidden transport realizations shown in (a).
}}
   \label{fig:fingerprint}  
\end{figure*}

\begin{figure}
  \includegraphics[trim=0 2mm 0 4mm,clip,width=1\linewidth]{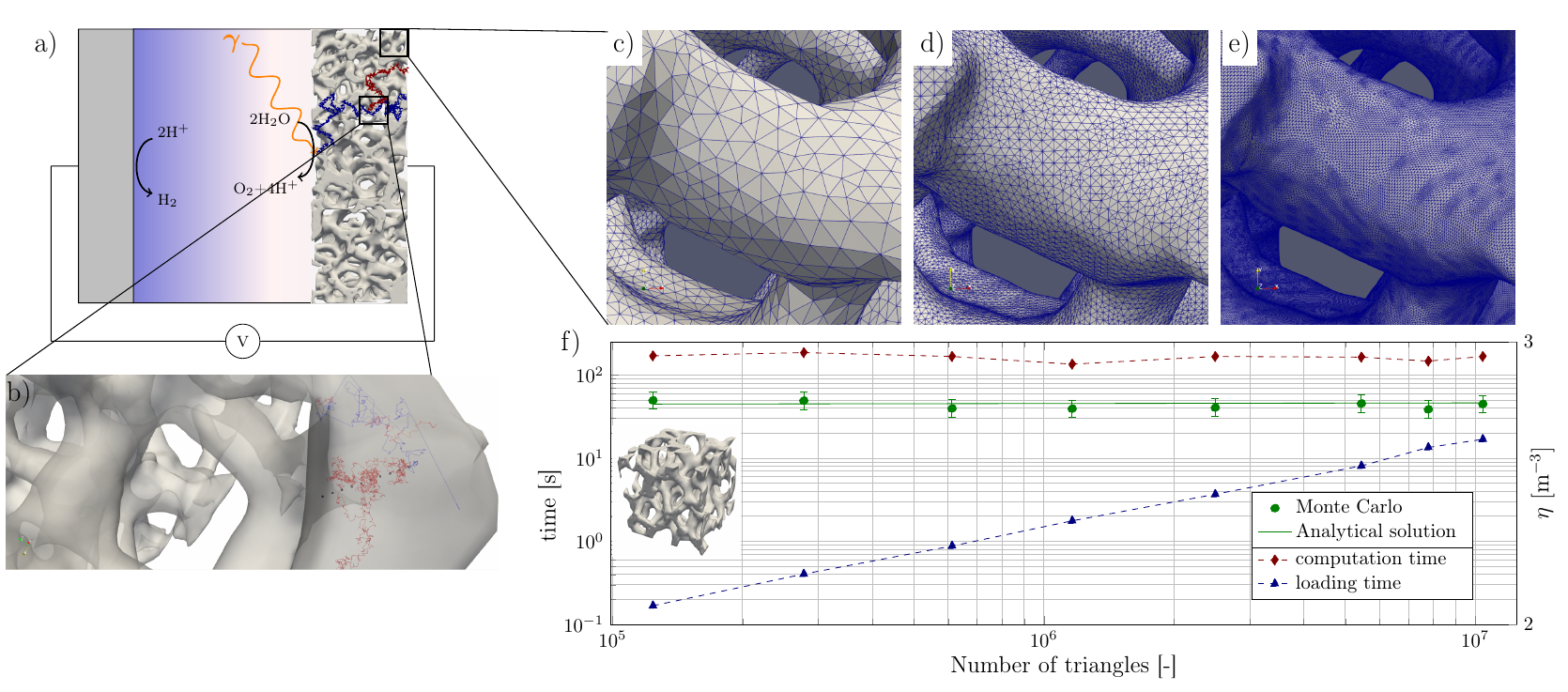}
  \caption{
    \scriptsize{a) Photocatalytic water splitting device involving charge carrier drift-diffusion transport.
    Sampled paths: main path in blue, drift sub-path in red.
    b) Zoom of the porous photoanode.
    c-e) Three different levels of refinement (\textit{i.e.} number of
    triangles) in the geometric description of the semiconducting
    crystal composing the photoanode, whose morphology is a key issue in
    attaining optimal efficiency.
    f) (\textcolor{black!50!green}{green}/right y-axis)
    Maruyama Monte-Carlo path-sampling of the propagation structure $(\boldsymbol{\gamma},\boldsymbol{\gamma}_\mathbf{t})$ compared
    to an analytical solution (see Appendix). Statistical confidence intervals provide a standard $1/\sqrt{N}$ behavior, for the sampling of $N$ joint histories. Error bars correspond to a one-$\sigma$ standard deviation. 
    (\textcolor{black!50!red}{red}) Computation time.
    (\textcolor{black!50!blue}{blue}) Geometry loading time.
  }}
  \label{fig:complex}
\end{figure}

Here we show that causality is not lost; it is hidden in \emph{branching histories}.
Restoring the hidden transport histories as explicit dynamical degrees of freedom lifts the effective dynamics to a joint history space in which information propagates locally and sequentially.
This reveals a common propagation structure shared across both kinetic and macroscopic descriptions.
Physics has repeatedly enlarged the notion of physical history, from configurations to trajectories and from trajectories to paths \cite{Wiener_1921,Onsager_1953,Gallavotti_1995,Garrahan_2007,Fodor_2016,Lecomte_2005}.
Self-consistent transport requires a further step: the elementary causal object is no longer an isolated path, but a branching history.
This joint history provides a direct and measurable organization of information flow, while remaining interpretable in complex geometries (Fig.~\ref{fig:complex}).
The physical consequence is striking: \emph{observable equivalence does not imply causal equivalence}.
A localized change in morphology can reorganize the ensemble of histories connecting a physical source to a probe while leaving the macroscopic response unchanged.
We demonstrate this by comparing two geometries admitting the same observable field.
Although the observables are strictly identical, the underlying causal fingerprints are remarkably different (Fig. \ref{fig:fingerprint}).
The source-to-observable map therefore reveals how morphology reorganizes the histories carrying an observable response, exposing information that is discarded by the observable field itself.

\begin{figure}[H]
  \centering
  \includegraphics[trim=1mm 2mm 5mm 1mm,clip,width=1\linewidth]{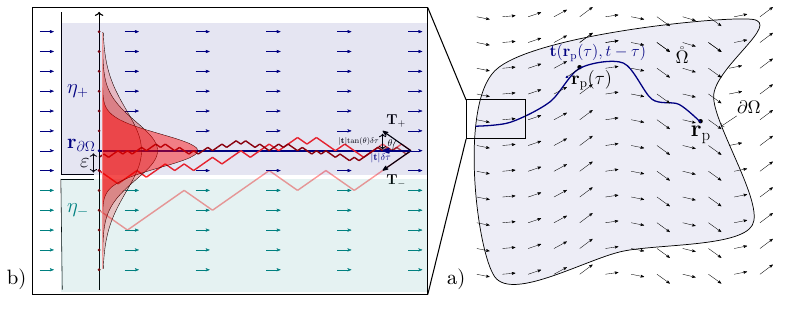}
  \caption{
   \scriptsize{
    a) Causal ballistic history traced for a prescribed transport field $\mathbf{t}$.
    b) Zoom close to the boundary, where $\eta(y>-\varepsilon)=\eta_+$ or
    $\eta(y<-\varepsilon)=\eta_-$ and $\mathbf{t}$ is uniform.
    [\textcolor{black!50!blue}{blue}] $\ds\mathbf{r}_\text{p}/\ds\tau=-\mathbf{t}$ straight-backwardly propagates up to the boundary where $\eta=\eta_+$.
    [\textcolor{black!50!red}{red}] Sample paths are broken lines since the transport field fluctuates around $-\mathbf{t}$.
    Apparent loss of direct reconstruction: hitting positions on the wall are spatially distributed, preventing the causal reconstruction of $\eta=\eta_+$
    (see Appendix).}}
   \label{fig:accroche}
\end{figure}

\textbf{Apparent loss of causality.} -- The difficulty is fundamentally related to the loss of a direct pathwise decomposition in the effective description of continuum transport when the transport field is not a fixed background field but is part of the dynamics and may itself depend on the observable \cite{McKean_1966}.
This can be seen in the simplest advective setting (Fig. \ref{fig:accroche}). 
When the transport field $\mathbf{t}$ is prescribed, a unique backward history $\ds \mathbf{r}_\text{p}/\ds\tau=-\mathbf{t}$ provides a direct causal map between a probe point $\mathbf{r}_\text{p}$ and the corresponding boundary location $\mathbf{r}_{\p\Omega}$. 
In that case, the observable $\eta$ can be directly read as the propagation of boundary information along a single history, so that $\eta(\mathbf{r})=\eta^{\partial\Omega}(\mathbf{r}_{\partial\Omega})$ in the purely advective and stationary regime.
If the transport field $\mathbf{t}$ depends on the current state along this history, this direct causal chain is lost. 
Indeed, in the simplest configuration of a fluctuating transport field centered along $-\mathbf{t}$, how can self-consistent transport still propagate information along well-defined paths?
Near a discontinuous boundary, the information is distributed across multiple incompatible histories.
The resulting mismatch of the \textit{joint} description makes the existence of a sharp deterministic transport response paradoxical and the emergence of standard propagation information structure seems elusive. 

In the most generic description of transport, the state vector $\boldsymbol{\eta}$ of the system evolves according to
\begin{equation}
\frac{\text{D}\boldsymbol{\eta}}{\text{D}t}=\cc{C}[\boldsymbol{\eta}]
\end{equation}
with a prescribed Initial Boundary Value $\boldsymbol{\eta}_\text{IBV}$.
 The transport acts through $\text{D}/\text{D}t=\p/\p t+\mathbf{t}\cdot\boldsymbol{\nabla}$, where $\mathbf{t}$ is the local transport field.
In kinetic theories, the mesoscopic transport of photons, neutrons, molecules, charges or globular clusters set by Boltzmann or Vlasov equations is characterized by the scalar phase-space distribution function $\eta(\mathbf{x},\mathbf{c},t)$.
 The local transport field stands then for $\mathbf{t}=(\mathbf{c},\mathbf{a})$ whereas the collision operator can either be written as $\cc{C}[\eta]=-(k_\text{a}+k_\text{d})\eta+\nu_\text{d}\int_{\bb{S}^2}\ds\boldsymbol{\omega'}~\varphi(\boldsymbol{\omega}|\boldsymbol{\omega'})\eta(\mathbf{x},|\mathbf{c}|\boldsymbol{\omega'},t)$ for linear transport theories such as photon scattering, or $\cc{C}[\eta]=\int_{\bb{R}^3}\ds\mathbf{c}_\star\int_{\bb{S}^2_+}\ds\mathbf{k}~g\sigma \eta(\mathbf{x},\mathbf{c'},t)\eta(\mathbf{x},\mathbf{c'}_\star,t)-\eta(\mathbf{x},\mathbf{c}_\star,t)\eta(\mathbf{x},\mathbf{c},t)$ for nonlinear transport theories such as gas kinetics.
In macroscopic transport theories, the state $\boldsymbol{\eta}(\mathbf{x},t)$ can either be characterized by scalar quantities such as charge density, chemical concentration or temperature when the evolution is set by Nernst--Planck or the heat equation, or vector fields such as the macroscopic velocity of a fluid when the evolution is set by Navier--Stokes or Burger's equations.
 The collision operator becomes $\cc{C}[\boldsymbol{\eta}]=D\nabla^2\eta$.

The value of the state at a given probe position $\mathbf{r}_\text{p}=(\mathbf{x})$ or $\mathbf{r}_\text{p}=(\mathbf{x},\mathbf{c})$ results then from the propagation of information paths connecting the considered effect $\eta(\mathbf{r}_\text{p},t)$ to initial and boundary physical sources:
\begin{equation}
\boldsymbol{\eta}(\mathbf{r}_\text{p},t)=\int_{\boldsymbol{\Gamma}}\cc{D}\mathds{P}[\boldsymbol{\gamma}]~\boldsymbol{\eta}_\text{IBV}(\boldsymbol{\gamma}[\cc{T}],t-\cc{T})
\end{equation}
The resulting summation over the physical sources reads as the path integral over the space $\boldsymbol{\Gamma}$ of information histories $\boldsymbol{\gamma}$ propagating backward in time from the observation event $\mathbf{r}_\text{p}$ at $\tau=\text{o}$ towards initial or boundary sources, until the stopping time $\tau=\cc{T}$.
Such causal paths propagate along $\ds \mathbf{r}_\text{p}/\ds \tau=-\mathbf{t}(\mathbf{r}_\text{p}(\tau),t-\tau)+\sqrt{2D}\boldsymbol{\xi}(\tau)+\boldsymbol{\cc{J}}(\mathbf{r}_\text{p}(\tau))\dot{\cc{N}}(\tau)$ given the standard gaussian white noise $\boldsymbol{\xi}$ and the jump process $\cc{N}$.
In kinetic theories $D=0$ and $|\boldsymbol{\cc{J}}|>0$ whereas in macroscopic transport models, $D>0$ and $|\boldsymbol{\cc{J}}|=0$.
 When $D=0$, $|\boldsymbol{\cc{J}}|=0$ and the state vector is a scalar function, the path-space measure refers to the deterministic Dirac measure
$\cc{D}\mathds{P}[\boldsymbol{\gamma}]\propto\cc{D}\boldsymbol{\gamma}~\delta(\dot{\boldsymbol{\gamma}}+\mathbf{t}(\boldsymbol{\gamma}))$ in free space.
 When $D>0$ and $|\boldsymbol{\cc{J}}|=0$, the path-space measure refers to the Wiener--Onsager--Machlup measure, which is explicit in free space: $\cc{D}\mathds{P}[\boldsymbol{\gamma}]\propto\cc{D}\boldsymbol{\gamma}~\text{exp}\left\{-\int_\text{o}^\cc{T}\frac{|\dot{\boldsymbol{\gamma}}+\mathbf{t}(\boldsymbol{\gamma})|^2}{4D}-\frac{1}{2}\boldsymbol{\nabla}\cdot\mathbf{t}(\boldsymbol{\gamma})\right\}$. 
 
\begin{figure}[H] 
  \centering
  \includegraphics[clip,width=1\linewidth]{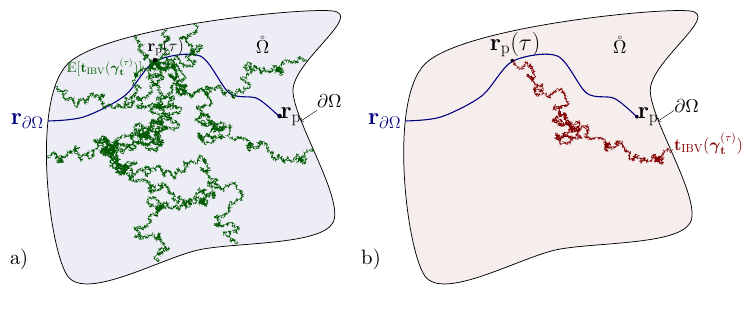}
  \caption{
   \scriptsize{
   a) effective description of self-consistent transport in the situation of Fig.~\ref{fig:accroche}: the main history must reconstruct the transport field histories at each location, effectively nesting an entire transport field path-space into the main trajectory.
   b) Joint description revealing the direct organization of information flow in self-consistent transport.}}
   \label{fig:McKY}  
\end{figure}

 When the transport field is self-consistently described by its own sub-model,  its value is obtained by propagating information from its own initial and boundary sources. 
 The transport field possesses therefore its own causal histories, which remain hidden in the \textit{effective} description.
Then, 
\begin{equation}
\mathbf{t}(\mathbf{r}_\text{p}(\tau),t-\tau)=\int_{\boldsymbol{\Gamma}_\mathbf{t}}\cc{D}\mathds{P}[\boldsymbol{\gamma}_\mathbf{t}^{(\tau)}]~\mathbf{t}_\text{IBV}(\boldsymbol{\gamma}_\mathbf{t}^{(\tau)}[\cc{T}])\label{eq:t}
\end{equation}
The path-space measure $\cc{D}\mathds{P}[\boldsymbol{\gamma}]$ therefore stands as an effective measure since the causal integration over transport field histories is then hidden.
 In this view, this effective transport measure appears to be explicitly nonlinear in the transport field and nonlocal, from which follows the loss of local information organization.
Since this measure depends on all transport histories, it implicitly contains the complete statistics of the transport-field measure.
Consequently, every history implicitly carries the statistics of the transport histories that determine its local propagation, leading to an apparent hierarchy of nested history spaces (Fig.~\ref{fig:McKY} a), known as McKean picture of self-consistent transport \cite{McKean_1966,Talay_2020,Calderoni_1983,Lejay_2020,Rioux_2022,Sugimoto_2024}. 
This loss of causal transparency can already be seen in the simplest confined advective setting (Fig. \ref{fig:accroche}). 
Several approaches have nevertheless intended to bypass this loss of information propagation. 
Transport signatures $\mathbf{t}\cdot\boldsymbol{\nabla}$ are interpreted as effective sources along the history of the transported state or as spectral interactions \cite{Labordere_2019,Nguwi_2023,Busnello_1999,Bhattacharya_2003,Ossiander_2005,LeJan_1997,Acebron_2013,Floriani_2008}. 
These approaches typically rely on Fourier-space formulations or Malliavin stochastic calculus \cite{Fournie_1999}, which are incompatible with confined and geometrically complex domains that arise in electrochemical devices (Fig.~\ref{fig:complex}), porous media, or biological ones. \\
 
 \textbf{Hidden transport information.} -- We show now that this hierarchy is entirely representational. 
 The nesting disappears once the hidden transport histories are restored as explicit dynamical variables. 
 Despite the paradox depicted in Fig. \ref{fig:accroche}, it is possible to restore local information propagation along well-defined  histories.
 Indeed, the spurious intuition leading to this paradox lies in the inconsistency of the discrete time picture.
 In the continuous time limit, propagating information along paths nested with complete history spaces $\boldsymbol{\Gamma}_\mathbf{t}$ \eqref{eq:t} of $\mathbf{t}$ or propagating information along paths coupled with single realizations of $\boldsymbol{\gamma}_\mathbf{t}^{(\tau)}$ is strictly equivalent (see Appendix).
Hence, causal continuity emerges with the natural propagation measure over the corresponding joint space:
 \begin{equation}
\boldsymbol{\eta}(\mathbf{r}_\text{p},t)=\int_{\boldsymbol{\Gamma}_\text{joint}\times\boldsymbol{\Gamma}_\mathbf{t}}\hspace{-0.4cm}\cc{D}\mathds{P}_\text{joint}[\boldsymbol{\gamma}_\text{joint},\boldsymbol{\gamma}_\mathbf{t}]~\boldsymbol{\eta}_\text{IBV}(\boldsymbol{\gamma}_\text{joint}[\cc{T}])
\end{equation}
As shown in the Appendix, both descriptions generate identical observable statistics, implying that the \textit{effective} measure is recovered as the observable marginal of the joint measure:
\begin{equation}
\cc{D}\mathds{P}[\boldsymbol{\gamma}]=\int_{\boldsymbol{\Gamma}_\mathbf{t}}\cc{D}\mathds{P}_\text{joint}[\boldsymbol{\gamma},\boldsymbol{\gamma}_\mathbf{t}]
\end{equation}
 
 In so doing, rather than describing transport through the effective marginalized measure, we lift the dynamics to the natural joint path measure of the transported state and the transport field.
 This lifting has a simple physical interpretation. 
 The transport trajectory no longer reconstructs the whole history space of the transport field. 
 Instead, it evolves conditionally on the knowledge of a complete transport history. 
 Every value of the transport field encountered along the trajectory is inherited from the same underlying realization. 
 Averaging over transport histories is performed only after the conditional transport problem has been solved.
 The fundamental object is therefore no longer an individual causal history $\boldsymbol{\gamma}$ but the joint history $(\boldsymbol{\gamma},\boldsymbol{\gamma}_\mathbf{t})$ formed by the transported trajectory and the transport-field trajectory. 
Thus, the effective description can be conceptualized as the marginalization procedure over hidden transport field histories which are latent dynamical variables. 
The resulting dynamics becomes non-local and generally non-Markovian \cite{Mori_1965,Zwanzig_1973}.
Introducing these latent variables explicitly lifts the problem to an enlarged path space where trajectories evolve conditionally on a single realization of the transport field history.

The auxiliary trajectory $\boldsymbol{\gamma}_\mathbf{t}^{(\tau)}$ appearing in the path integral representation of the local transport field should not be confused with the hidden history entering the joint path measure. 
The latter denotes the collection of realizations along the observable history: $\boldsymbol{\gamma}_\mathbf{t}=\{\boldsymbol{\gamma}_\mathbf{t}^{(\tau)}\}_\tau$ (Fig. \ref{fig:McKY}). 
The existence of a joint measure therefore lifts the hidden transport field histories to the level of complete histories. 
Apparent nonlocality emerges only after marginalization of hidden transport field histories. 
Reintroducing these latent variables restores a local conditional dynamics in an enlarged path space and reveals that transport information is carried locally and sequentially by a single realization of the hidden transport process.
The effective description appears to be the marginal dynamics of this natural joint measure, explaining why causal propagation emerges from the lifted dynamics.

\textbf{Causal fingerprint of morphology.} -- The joint-histories reveals the physical level of information organization that is hidden in the usual observable description. 
We next show that this signature can distinguish geometries that are indistinguishable at the level of the observable state (Fig. \ref{fig:fingerprint} (a)).
We demonstrate that a local change in morphology reorganizes entirely the causal histories connecting physical sources to a probe while the observable state remains unchanged. 
Let us consider the stationary transport $\mathbf{t}\cdot\boldsymbol{\nabla}\eta=D\nabla^2\eta-(\mu q/\varepsilon)\eta$ of charges concentration $\eta$ hierarchicaly coupled to the self-consistent model $\boldsymbol{\nabla}\cdot\mathbf{t}=\mu q/\varepsilon$ of the transport-field $\mathbf{t}$ in two domains, $\Omega_1\equiv[\text{o};l]^2$ and $\Omega_2\equiv[\text{o};l]^2\backslash([l-\delta;l]\times[\text{o};\delta])$, sharing the same external perimeter $|\p\Omega|$ and observation point $\mathbf r_{\rm p}$, but differing by a localized modification of the boundary. 
Boundary conditions enforce that the observable state $\eta(\mathbf{r}_\text{p})$ of the system is strictly identical in both geometries. 
From the same probe point, $N=10^4$ backward joint histories $(\boldsymbol{\gamma},\boldsymbol{\gamma}_\mathbf{t})$ are sampled independently for each morphology until reaching a boundary source. 
Each realization $i\in\llbracket1;N\rrbracket$ carries a contribution weight $w_i$ to the observable.
Despite the exact equivalence of the observable response, the corresponding history ensembles are not equivalent and this difference cannot be interpreted as a difference already present in the observable state which is held fixed. 
The morphological perturbation therefore leaves a measurable imprint on the organization of the physical information propagated toward $\mathbf{r}_\text{p}$.
This reorganization can be resolved at several complementary levels as shown in the panels (b)--(d) of Fig. \ref{fig:fingerprint}. The boundary-hit distribution describes the geometric accessibility of the sources. The contribution-weighted distribution $W_\eta(s_{\p\Omega})$ instead describes how the observable response is distributed over those sources. 
These quantities need not coincide: frequently visited boundary regions may carry little of the response, whereas comparatively rare histories may contribute disproportionately. The same distinction appears along the histories themselves. Weighting the sampled local transport states by $w_i$ identifies the portions of the hidden transport environment that effectively participate in the probe response. 
Conditioning these histories on the propagated observable further resolves correlations between the observed state and the local environments through which its contribution has been assembled.

We refer to this complete source-to-probe organization as the causal fingerprint $\cc{F}$ of the morphology. 
Importantly, this fingerprint reveals the physical redistribution of information in the joint histories of the system.
The two geometries satisfy
\begin{equation}
\begin{split}
\eta_{\Omega_1}(\mathbf r_{\rm p})&=\eta_{\Omega_2}(\mathbf r_{\rm p})\\
\mathcal F_{\Omega_1}(\text{source}\rightarrow\text{probe})&\neq\mathcal F_{\Omega_2}(\text{source}\rightarrow\text{probe})
\end{split}
\end{equation}
The distinction is quantitative. 
For any pair of normalized causal distributions $p_{\Omega_1}(s_{\p\Omega})$ and $p_{\Omega_2}(s_{\p\Omega})$ defined over the same curvilinear coordinate $s_{\p\Omega}$, their causal distinguishability can be quantified by the Kullback--Leibler divergence defined as 
\begin{equation}
\cc{D}_\text{KL}(p_{\Omega_1}||p_{\Omega_2})=\int_{|\p\Omega|} \hspace{-0.15cm}\ds s_{\p\Omega}~p_{\Omega_1}(s_{\p\Omega})\text{ln}\left(\frac{p_{\Omega_1}(s_{\p\Omega})}{p_{\Omega_2}(s_{\p\Omega})}\right)
\end{equation}
This configuration displays $\cc{D}_\text{KL}(W_\eta^{\Omega_1}||W_\eta^{\Omega_2})=1.2$. 
In other words, when sampled according to $W_\eta^{\Omega_1}$, the geometric mean likelihood ratio is $e^{1.2}\approx 3.3$. 
A wide range of complementary distributions can depict the causal fingerprint. 
Indeed, the fingerprint is not a single observable quantity, but a family of distributions obtained by projecting, weighting, and conditioning the ensemble of joint histories. 
For instance, the contribution-weighted spatial distribution of the hidden transport-field histories $W_\mathbf{t}(\Delta\mathbf{x}_\mathbf{t})$ realted to their boundary positions $\Delta\mathbf{x}_\mathbf{t}=\mathbf{x}_\mathbf{t}-\mathbf{x}_\mathbf{t}^{\p\Omega}$ (Fig.~\ref{fig:fingerprint}(b)) reveals which regions of the transport environment participate most strongly in carrying the probe response.
In the same vein, the contribution-weighted transport-history positions conditioned on the observable state $W_\mathbf{t}(\mathbf{x}_\mathbf{t}|\eta)$ shown in Fig.~\ref{fig:fingerprint}(d)  identifies the regions of the hidden transport environment associated with histories carrying a selected range of observable values. 
Beyond these spatial measures, the causal fingerprint can also be characterized through history-level statistics such as path length, exit time, number of transport steps, accumulated weights, or concentration of the contribution among histories. 

A local morphological perturbation can therefore leave the observable unchanged while producing a measurable redistribution of causal information.
The comparison reveals how the morphology reorganizes the local transport environments actually sampled by the joint histories.
 Two morphologies can thus be observationally identical while remaining causally distinguishable. 
In this sense, the causal fingerprint makes explicit physical information that is present in the self-consistent dynamics but is hidden by the effective description at the level of the observable field. 

This distinguishable organization of information flow across joint histories becomes particularly relevant in complex geometries, where geometric refinement can strongly alter the accessibility and connectivity of causal histories. We now illustrate in Fig. \ref{fig:complex} this property in self-consistent charge transport in a nanoporous semiconductor relevant to artificial photosynthesis (see Appendix). 
The resulting sampling procedure retains several standard advantages of Monte Carlo approaches.
Because the sampling is mesh-free, naturally parallelizable, and pointwise, computation times remain insensitive to geometric refinement.\\

\begin{figure}[H] 
  \centering
  \includegraphics[clip,width=1\linewidth]{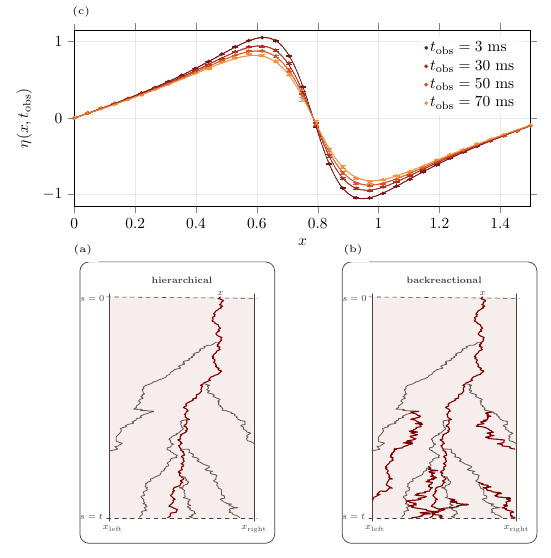}
  \caption{
   \scriptsize{Burger's dynamics
$\p_t \eta+\eta\p_x\eta=\nu\p_x^2\eta$ provides a minimal backreaction self-consistent transport benchmark in which the transport field is itself the observable field. 
The corresponding causal signature is a binary branching tree (b) in contrast with the joint history of hierarchical transport.
}}
   \label{fig:Burger}  
\end{figure}

\textbf{Self-consistent transport with backreaction.} -- We showed that the causal structure of self-consistent transport is naturally encoded in a joint path measure. 
In the case of an observable history $\boldsymbol{\gamma}$ whose dynamics hierarchically depends on an autonomous transport-field history $\boldsymbol{\gamma}_\mathbf{t}$, the joint measure can formally be written as $\cc{D}\mathds{P}_\text{joint}[\boldsymbol{\gamma},\boldsymbol{\gamma}_\mathbf{t}]=\cc{D}\mathds{P}[\boldsymbol{\gamma}_\mathbf{t}]\cc{D}\mathds{P}[\boldsymbol{\gamma}|\boldsymbol{\gamma}_\mathbf{t}]$. 
When the transport field depends back on the observable dynamics, the correct object is no longer a single path conditioned on a fixed transport field history, but a rooted branching history. 
Backreaction turns a single history into a branching history tree, with each descendant branch generated from the instantaneous state of its parent branch (Fig.~\ref{fig:Burger} (b)). 
In this recursive case the same structure holds locally at each branching event, with the global joint law generated as a product of local conditional kernels along the tree. 
Importantly, the propagation remains local and sequential while the effective observable measure is recovered by marginalization.
A single causal path is therefore promoted to a branching history tree, in which each branch carries the state inherited from its parent and can in turn generate new descendants. 
The resulting tree is the causal signature of self-consistency, making explicit the hierarchy of successive feedback events that is hidden by the effective, marginalized transport description.
As shown in Fig. \ref{fig:Burger} (c), these trees operationnally solve the minimal backreaction benchmark of the Bruger's dynamics $\p_t\eta+\eta\p_x\eta=\nu\p_x^2\eta$ (see Appendix).

\textbf{Conclusion.} -- Causality is not lost in self-consistent transport; it is hidden in joint histories.
These information paths are hidden in the usual description, obscuring the paths by which a physical source influences an observable.
When these histories are made explicit, information flow appears as a branching history in which backreactions are sequential.
This reveals a physical organization that cannot be seen from the observable field alone.
A local change in morphology can completely reorganize the histories connecting sources to an observation while leaving the observed state unchanged.
The observable therefore does not contain the whole information about how it is produced: two systems can be observationally identical while having different causal organizations.
The causal fingerprint provides a way to reveal this hidden organization.
It shows where the information reaching an observation comes from and through which transport histories it has propagated.
More generally, our results suggest that in self-consistent transport, causality is not carried by an isolated path, but by a branching history.
Making this hidden structure visible opens a way to study how geometry and interactions organize the propagation of physical information beyond what is revealed by observables alone.

\begin{center}
***
\end{center}
  
The authors thank \DIFdelbegin M\'esoStar 
(\url{www.meso-star.com}) for their support in producing Fig.~\ref{fig:complex}.
This work was supported by the International Research Center "Innovation Transportation and Production Systems" of the I-SITE CAP 20-25 (ANR-16-IDEX-0001) and the MCMET project (ANR-23-CE46-0002) of the French National Research Agency (ANR).\\


\textit{Data availability:} The datasets used and/or analysed during the current study available from the corresponding author on reasonable request.

\bibliographystyle{apsrev4-2}
\bibliography{biblio_merge}
 
\textbf{Appendix.} -- \textit{Equivalence between the effective measure and the marginalized joint measure.} For the sake of simplicity, consider the joint process without jump process: $\ds\mathbf{r}_\text{p,joint}/\ds\tau=-\mathbf{t}_\text{IBV}[\boldsymbol{\gamma}_\mathbf{t}^\tau]+\sqrt{2D}\boldsymbol{\xi}(\tau)$ given $\mathbf{r}_\text{p,joint}(\text{o})=\mathbf{r}_\text{p}$. 
The corresponding joint path measure is denoted 
$\cc{D}\mathds{P}_\text{joint}[\boldsymbol{\gamma}_\text{joint},\boldsymbol{\gamma}_\mathbf {t}]$ 
where $\boldsymbol{\gamma}_\mathbf{t}=\{\boldsymbol{\gamma}_\mathbf{t}^{(\tau)}\}_\tau$.
Then, $\boldsymbol{\eta}(\mathbf{r}_\text{p}(\tau),t-\tau)$ follows
$\ds\boldsymbol{\eta}=\left(\p_\tau\boldsymbol{\eta}-\mathbf{t}_\text{IBV}\cdot\boldsymbol{\nabla}\boldsymbol{\eta}+D\boldsymbol{\nabla}^2\boldsymbol{\eta}\right)\ds \tau+\sqrt{2D}\boldsymbol{\nabla}\boldsymbol{\eta}\cdot \ds\boldsymbol{\mathfrak{W}}_\tau
$ by Itô's lemma. 
Taking the conditional expectation with respect to $\mathbf{r}_\text{p}$ leads then to $\p_t\boldsymbol{\eta}+\mathbf{t}\cdot\boldsymbol{\nabla}\boldsymbol{\eta}=D\boldsymbol{\nabla}^2\boldsymbol{\eta}$ given
$\mathbf{t}(\mathbf{r}_\text{p},t)=\int_{\boldsymbol{\Gamma}_\mathbf{t}}\cc{D}\mathds{P}[\boldsymbol{\gamma}_\mathbf{t}^{(\tau)}]~\mathbf{t}_\text{IBV}(\boldsymbol{\gamma}_\mathbf{t}^{(\tau)}[\cc{T}])
$. 
This is precisely the backward Kolmogorov equation associated with the effective process $\ds\mathbf{r}_\text{p,eff}/\ds\tau=-\mathbf{t}+\sqrt{2D}\boldsymbol{\xi}(\tau)$ whose path measure is denoted $\cc{D}\mathds{P}_\text{eff}$.
Since both representations satisfy the same backward equation with the same IBV uniqueness implies
$\int_{\boldsymbol{\Gamma}_\text{joint}\times\boldsymbol{\Gamma}_\mathbf{t}}\cc{D}\mathds{P}_\text{joint}[\boldsymbol{\gamma}_\text{joint},\boldsymbol{\gamma}_\mathbf{t}]\boldsymbol{\eta}_\text{IBV}(\boldsymbol{\gamma}_\text{joint}[\cc{T}],t-\cc{T})=\int_{\boldsymbol{\Gamma}_\text{eff}}\cc{D}\mathds{P}_\text{eff}[\boldsymbol{\gamma}_\text{eff}]\boldsymbol{\eta}_\text{IBV}(\boldsymbol{\gamma}_\text{eff}[\cc{T}],t-\cc{T})
$
For every admissible IBVP, probe position and time, both
representations therefore yield the same observable. 
Introducing the marginal measure $\cc{D}\mathds{P}_\text{marg}[\boldsymbol{\gamma}_\text{joint}]$, 
one concludes $\cc{D}\mathds{P}_\text{eff}[\boldsymbol{\gamma}]=\cc{D}\mathds{P}_\text{marg}[\boldsymbol{\gamma}]$. Hence the effective measure coincides with the marginal measure of the joint histories.

\textit{Continuous-time limit for Fig.~\ref{fig:accroche}.} The random transport field $\mathbf{T}$ is defined as
$\mathbf{T}=|\mathbf{t}|(-1,(1-2\cc{B}(\frac{1}{2}))\text{tan}(\theta))$,
where $\cc{B}(\frac{1}{2})$ is the Bernoulli variable of probability $1/2$.
$\mathbf{T}$ outcomes are $\mathbf{T}_+=|\mathbf{t}|(-1,\text{tan}(\theta))$ and $\mathbf{T}_-=(-1,-\text{tan}(\theta))$, and $\mathds{E}[\mathbf{T}]=|\mathbf{t}|(-1,0)=-\mathbf{t}$. 
Starting from $\mathbf{r}_\text{p}=(l,y)$, the first hitting locations of the purely advective paths to the left-sided boundary wall are given by $\mathbf{r}_\text{p,joint}(N\delta \tau)=\mathbf{r}_\text{p}+\sum_{i=0}^{N-1}\mathbf{T}_i\delta \tau$ with $N=l/(|\mathbf{t}|\delta \tau)$. 
Its horizontal component is null by construction. 
Therefore, only first passage heights $y_\text{p,joint}(N\delta \tau)$ are
distributed and their realizations can take values in
$\{k|\mathbf{t}|\delta \tau~\text{tan}(\theta)|k\in\bb{Z}\}$. 
As the result of successive Bernoulli trials, the first passage height
distribution follows a binomial law behaving as a gaussian distribution with mean $\mathds{E}[y_\text{p,joint}(l/|\mathbf{t}|\delta \tau)]=y$ and variance $\mathds{V}[y_\text{p,joint}(l/|\mathbf{t}|\delta \tau)]=l|\mathbf{t}|\delta \tau~\text{tan}^2(\theta)$ vanishing as $\delta \tau\to0$.
In the situation depicted in Fig.~\ref{fig:accroche} the exact analytical
solution is $\eta(\mathbf{r}_\text{p})=\eta^{\p\Omega}(y)=\mathds{1}_{\{y>-\varepsilon\}}\eta_++\mathds{1}_{\{y\leq-\varepsilon\}}\eta_-$.
One has $\lim_{\delta \tau\to0}\mathds{E}\left[\eta^{\p\Omega}(y_\text{p,joint}(l/|\mathbf{t}|\delta \tau))\right]=\eta(\mathbf{r}_\text{p})$ and $\lim_{\delta \tau\to0}\mathds{V}\left[\eta^{\p\Omega}(y_\text{p,joint}(l/|\mathbf{t}|\delta \tau))\right]=0$.
In the continuous-time limit the backward joint information paths $\ds\mathbf{r}_\text{p,joint}/\ds\tau=\mathbf{T}$ therefore provide the exact solution of the boundary-value problem.\\

\textit{Benchmark for Fig.~\ref{fig:complex}.} We solve the semiconductor electron density $\eta$ submitted to $\mathbf{t}\cdot\boldsymbol{\nabla}\eta=D\nabla^2\eta$ and coupled to the external model of the electrostatic transport-field $\boldsymbol{\nabla}\cdot\mathbf{t}=0$.
In order to compare statistical estimations to an analytic solution of the
coupled confined system, Dirichlet boundary values coincide to a particular solution of the system: $\eta^{\p\Omega}(\mathbf{r}_{\p\Omega})=\sqrt{\frac{D}{\mu}}\left(\text{erf}\left(x\sqrt{\frac{\mu}{2D}}\right)+\text{erf}\left(y\sqrt{\frac{\mu}{2D}}\right)\right)+\sqrt{\frac{D}{\mu}}~\text{erfi}\left(z\sqrt{\frac{\mu}{D}}\right)$ and $\mathbf{t}^{\p\Omega}(\mathbf{r}_{\p\Omega})=(x,y,-2z)$.
Maruyama-Monte-Carlo numerical estimation of the charge density $\eta$ is evaluated at location ($x=y=1.05$, $z=0.9$) within the porous semiconductor by sampling $10^{3}$ information paths $(\boldsymbol{\gamma},\boldsymbol{\gamma}_\mathbf{t})$ and considering $\delta \tau=4\times 10^{-4}$ s, $\mu=1$
m$^2$.V$^{-1}$.s$^{-1}$ and $D=1$ m$^2$.s$^{-1}$. 
Results are plotted against the analytic solution for several geometric refinements.\\

\textit{Benchmark for Fig.\ref{fig:fingerprint}.} The prescribed charge density is $q(\mathbf{r})=\frac{4\varepsilon D}{\mu}\sum_{m,\sigma}\xi_m^{(\sigma)}|\mathbf{k}_m^{(\sigma)}|^2\text{cos}(\Psi_m^{(\sigma)}(\mathbf{r}))$, given $\Psi_m^{(\sigma)}(\mathbf{r})=\mathbf{k}_m^{(\sigma)}\cdot(2\mathbf{r}-\boldsymbol{1})+\theta_m$. 
Dirichlet boundary conditions $\eta(\mathbf{r})=\eta_\text{o}\text{exp}\{\sum_{m,\sigma}\xi_m^{(\sigma)}\text{cos}(\Psi_m^{(\sigma)}(\mathbf{r}))\}$ and $\mathbf{t}(\mathbf{r})=2D\sum_{m,\sigma}\xi_m^{(\sigma)}\mathbf{k}_m^{(\sigma)}\text{sin}(\Psi_m^{(\sigma)}(\mathbf{r}))$ are prescribed for $\mathbf{r}\in\p\Omega$ and coincides with a particular solution of the system. 
The modes are divided into three spatial scales $\sigma\in\{\mathrm{L},\mathrm{M},\mathrm{S}\}$, with \(N_{\mathrm L}=2\), \(N_{\mathrm M}=5\), and \(N_{\mathrm S}=5\) modes. 
For each mode, the wave-vector 
magnitude is sampled uniformly within the scale-dependent interval $|\mathbf k_m^{(\mathrm L)}|\in[\frac{0.6\pi}{l};\frac{1.2\pi}{l}]$, $|\mathbf k_m^{(\mathrm M)}|\in[\frac{1.4\pi}{l};\frac{2.4\pi}{l}]$, and $|\mathbf k_m^{(\mathrm S)}|\in[\frac{2.8\pi}{l},\frac{4.2\pi}{l}]$. 
Thus, $\mathbf k_m^{(\sigma)}=|\mathbf k_m^{(\sigma)}|(\cos\alpha_m^{(\sigma)},\sin\alpha_m^{(\sigma)})$ with \(\alpha_m^{(\sigma)}\sim\mathcal U([0;2\pi]\).
The phase follows $\theta_m^{(\sigma)}\sim\mathcal U([0;2\pi]$.
The mode amplitudes are generated from $\xi_m^{(\sigma)}=\Phi_{\mathrm{scale}}\,A_\sigma\,U_m^{(\sigma)}$ with $U_m^{(\sigma)}\sim\mathcal U([0.7;1.3])$,  $A_{\mathrm L}=0.4$, $A_{\mathrm M}=0.25$, $A_{\mathrm S}=0.12$, and $\Phi_{\mathrm{scale}}=2\times10^{-2}$.
Since the electrostatic transport field satisfies $\boldsymbol{\nabla}^2\mathbf t=\frac{\mu\boldsymbol{\nabla}q}{\varepsilon}$, it results from the causal propagation of boundary values $\mathbf{t}^{\p\Omega}(\boldsymbol{\gamma}_\mathbf{t})$ and volumic sources 
$\mathbf{t}^\text{S}(\boldsymbol{\gamma}_\mathbf{t})=-\frac{\mu}{\varepsilon}\int_\text{o}^{\cc{T}_{\p\Omega}}\ds\zeta~\boldsymbol{\nabla}q(\boldsymbol{\gamma}_\mathbf{t}(\zeta))$.
Transport-field paths $\boldsymbol{\gamma}_\mathbf{t}$ are described by
$\ds\mathbf{r}_\text{p}(\tau)/\ds\tau=\sqrt{2}\boldsymbol{\xi}(\tau)$.
The observable concentration results from the attenuated/amplified propagation of its boundary values:
$\eta(\mathbf{r})=\int_{\boldsymbol{\Gamma}}\cc{D}\mathds{P}[\boldsymbol{\gamma}]~\eta^{\p\Omega}\left(\boldsymbol{\gamma}(\cc{T}_{\p\Omega})\right)\text{exp}\left\{\frac{\mu}{\varepsilon}\int_\text{o}^{\cc{T}_{\p\Omega}}\ds\zeta~q(\boldsymbol{\gamma}(\zeta))\right\}
$ 
with the path $\boldsymbol{\gamma}$ described by 
$\ds\mathbf{r}_\text{p,joint}(\tau)/\ds\tau=[\mathbf{t}^{\p\Omega}(\boldsymbol{\gamma}_\mathbf{t}^{(\tau)})+\mathbf{t}^\text{S}(\boldsymbol{\gamma}_\mathbf{t}^{(\tau)})]+\sqrt{2D}\ds\boldsymbol{\xi}(\tau)$.
In the benchmark, $D=1\times 10^{-1}$ [m$^2$.s$^{-1}$], $l=5\times 10^{-1}$ [m], $\delta=1.87\times 10^{-1}$ [m], $\mu=1$ [m$^2$.s$^{-1}$.V$^{-1}$], and $\eta_\text{o}=1$ [m$^{-2}$] are considered.
Monte Carlo statistical estimations obtained by use of $N=1\times 10^4$ realizations of this branching process discretized with the Euler--Maruyama step $\delta \tau=1\times 10^{-4}$ [s] are compared to the analytical solution  for $\mathbf{r}=(3\times 10^{-1},2\times 10^{-1})\in\Omega_{1}\cap\Omega_2$.
As a result, $\eta_{\text{MC},\Omega_1}(\mathbf{r})=1.02366\pm 5.8\times 10^{-4}$ and $\eta_{\text{MC},\Omega_2}(\mathbf{r})=1.02349\pm 3.3\times 10^{-4}$. 
Both Monte Carlo estimates converge with a $99.73\%$ confidence intervals to the analytical solution $\eta_\text{ana}=1.02341$ within \(0.3\%\).\\

\textit{Benchmark for Fig.\ref{fig:Burger}.} Consider within the confined domain $\Omega\equiv[x_\text{left}=0;x_\text{right}=1.5]$ [m] the viscous Burger dynamics $\p_t\eta(x,t)+\eta(x,t)\p_x\eta(x,t)=\nu\p_x^2\eta(r,t)$. 
We impose the boundary condition $\eta(x,t)=2\nu k\alpha(t)\text{sin}(kx)/(1+\alpha(t)\text{cos}(kx))$ with $\alpha(t)=ae^{-\nu k^2t}$ for $x\in\p\Omega$ and the initial condition $\eta(x,\text{o})=2\nu ka~\text{sin}(kx)/(1+a~\text{cos}(kx))$.
In the benchmark, $k=4$ [m$^{-1}$], $a=8\times 10^{-1}$ [.] and $\nu=1\times 10^{-1}$ [m$^2$.s$^{-1}$], are considered.
Monte Carlo statistical estimations obtained by use of $10^3$ branching paths sampled with $\delta\tau=t_\text{obs}/20$ are compared to the Cole--Hopf solution $\eta_\text{ana}(x,t)=2\nu k\alpha(t)\text{sin}(kx)/(1+\alpha(t)\text{cos}(kx))$. 

\end{document}